\documentclass[epj]{svjour}
\usepackage{graphicx}
\usepackage{amssymb}

\begin{document}

%%%%%%%%%%%%%%%%%%%%%%%%%%%%%%%%%%%%%%%%%%%%%%%%%%%%%%%%%%%%%

\title{Medium effects on $\phi$ decays to dilepton and kaon-antikaon 
pairs in relativistic heavy ion reactions}

\author{E. Santini, G. Burau, Amand F{\"a\ss}ler, C. Fuchs}
\institute{Institut f{\"u}r Theoretische Physik, 
Universit{\"a}t T{\"u}bingen, Auf der Morgenstelle 14, 
D-72076 T{\"u}bingen, Germany}

\date{Received: date / Revised version: date}

\authorrunning{E. Santini {\it et al.}}

\titlerunning{Medium effects on $\phi$ decays to dilepton and kaon-antikaon 
pairs ...}

%%%%%%%%%%%%%%%%%%%%%%%%%%%%%%%%%%%%%%%%%%%%%%%%%%%%%%%%%%%%%%

\abstract{
We consider the role of rescattering of secondary kaons on the dilepton 
branching ratio of the $\phi$ meson. In-medium mass modifications and 
broadening of kaons and $\phi$ mesons are taken into account. 
We find in the framework of a Bjorken scenario for the time evolution 
of the expanding fireball that the $\phi$ yield from dimuons is moderately 
or at least only slightly enhanced compared to that from kaon-antikaon pairs. 
The relation to experimental yields measured by the NA49, NA50 and CERES 
Collaborations at CERN SPS and the PHENIX Collaboration at RHIC is discussed.
 % end of PACS codes
} %end of abstract

%\begin{keyword}
%ultra-relativistic heavy ion collisions \sep
%$\phi$ meson production and decay \sep
%in-medium modification
%\end{keyword}

\maketitle

%%%%%%%%%%%%%%%%%%%%%%%%%%%%%%%%%%%%%%%%%%%%%%%%%%%%%%%%%%%%%%%

\section{Introduction}

Relativistic heavy ion (RHI) collisions present an unique opportunity 
to create nuclear matter at high density and temperature where hadron 
properties become different and the phase transition to quark matter 
with signature of the deconfinement and restoration of chiral symmetry 
is expected. Strangeness enhancement in relativistic nucleus-nucleus 
collisions compared to nucleon-nucleon collisions has been suggested 
as a possible signal for the formation of the quark gluon plasma (QGP) 
\cite{rafelski82}. The dominant production of $s\bar{s}$ pairs via 
gluon-gluon interaction in the plasma may result in an enhanced number 
of strange and multi-strange particles produced after hadronization. 
In particular, the production and decay of the $\phi$ meson have been 
long recognised as an important probe for the state of matter produced 
in RHI collisions \cite{rafelski82,shor85}. The possible enhancement 
of its yield has been suggested as an important signature of the formation 
of a QGP state \cite{shor85}, in which free $s\bar{s}$ pairs would coalesce 
to form $\phi$ mesons, whereas their production in $pp$ collisions is 
suppressed according to the Okubo-Zweig-Iizuka rule \cite{ozi}.

Phi meson production was measured in central Pb+Pb collisions at 158 $A$ GeV 
at the CERN SPS. The NA49 Collaboration identified the phi meson via the 
decay channel $\phi\rightarrow K^+K^-$ \cite{NA49}, while the NA50 
Collaboration identified it via the $\phi\rightarrow\mu^+\mu^-$ decay 
\cite{NA50}. It was found that the extracted number of phi mesons from the 
dimuon channel exceeded those number extracted from the $K^+K^-$ channel 
by a factor between two and four \cite{Rohrich:2001qi}.

It has been argued that this difference could be at least partially 
attributed to the fact that not all phi mesons can be reconstructed via 
invariant mass technique applied to $K^+K^-$ pairs. Due to the rescattering 
of the $K^+$ and $K^-$ in the medium, the reconstructed invariant mass could 
result outside the original phi meson peak in such a way that some 
$\phi\rightarrow K^+K^-$ events contribute to the experimental background 
rather than to the measured yields \cite{Johnson:1999fv}. The modification 
of the visible $\phi$ spectra due to kaon rescattering was first investigated 
within a basic RQMD model, showing that kaon rescattering can account for the 
observed rapidity distributions, but not for the differences in the $m_t$ 
spectra, neither the inverse slopes nor the relative yields, with a resulting 
suppression of the $\phi$ meson yield via the hadronic $K\bar{K}$ channel 
of 40-60\% \cite{Johnson:1999fv}. This effect has been further studied in 
the framework of a spherically expanding fireball model by including different 
possible changes (increasing/decreasing) of kaon masses in the hadronic medium 
\cite{Filip:2001st}, as well as in the framework of a multi-phase transport 
(AMPT) model by including density dependent in-medium modifications of kaons 
and $\phi$ mesons \cite{Pal:2002aw}. In none of these cases the large 
difference between the two SPS experiments could be explained. Such a large 
enhancement of four at low transverse mass from dimuon over the dikaon channel 
could only be explained in the AMPT model by an ad hoc increase of the phi 
meson number from the initial stage by about a factor of three, that might 
then suggest that other mechanisms, such as formation of colour ropes or 
quark-gluon plasma, are needed for $\phi$ meson production \cite{Pal:2002aw}.

To shed light on this so called ``$\phi$ puzzle'', two experimental 
collaborations planned to measure the $\phi$ yield simultaneously in both 
$\phi \rightarrow e^{+}e^{-} $ and $K\bar{K}$ channels at two different energy 
regimes, namely the CERES Collaboration at CERN SPS and the PHENIX 
Collaboration at the BNL Relativistic Heavy Ion Collider (RHIC). 
The CERES Collaboration measured $\phi$ meson production in central 
${\rm Pb+Au}$ collisions at $E_{lab}=158~A$ GeV. Recently reported data show 
that the yields and inverse slope parameters in both decay modes agree within 
errors. A yield in the $e^{+}e^{-}$ decay mode larger than 1.6 times the yield 
in the $K^+K^-$ channel is excluded at 95\% CL \cite{Adamova:2005jr}. 
The PHENIX Collaboration measured the $\phi$ yield in ${\rm Au+Au}$ and 
${\rm d+Au}$ collisions at $\sqrt{s_{\rm NN}}= 200~{\rm GeV}$ 
\cite{Nagle:2002ib,Adler:2004hv,Ozawa:2005zi,Kozlov:2005zy}. 
Whereas definitive results on the $\phi$ yield at mid-rapidity in the 
${\rm Au+Au}$ collisions as measured in the $K^+K^-$ channel have been 
already reported \cite{Adler:2004hv}, only preliminary results on the 
corresponding dilepton channel are available so far 
\cite{Nagle:2002ib,Ozawa:2005zi,Kozlov:2005zy}. According to recently reported 
preliminary production results, the temperatures measured in the two decay 
channels are in agreement within errors, while the yield in the $e^{+}e^{-}$ 
channel seems to be larger than in the $K^+K^-$ decay channel. However, the 
errors in the dielectron channel, both statistical and systematic, are still 
too large for a definite statement \cite{Kozlov:2005zy}. On the other side, 
recently reported preliminary data on the yields of both 
$\phi\rightarrow K^+K^-$ and  $\phi \rightarrow e^{+}e^{-}$ in ${\rm d+Au}$ 
collision show that the two data samples are consistent in both the total 
invariant yield and inverse slope \cite{Ozawa:2005zi}.

In heavy ion reactions at energies reached at RHIC, the medium after 
hadronization is baryon dilute but meson rich \cite{bravina02}. Since by far 
the most abundant particles are pions, one can speak about pion matter created 
in heavy ion collisions at RHIC energies. The kaon-pion gas was studied in 
detail within chiral perturbation theory (ChPT) in ref. \cite{dobado}, 
focusing thereby on the quark condensates. ChPT, proposed for the description 
of interactions of pseudo-scalar mesons at low energies, although being an 
adequate tool for studying the problem at low temperatures 
($T\lesssim M_{\pi }$), cannot be applied at typical temperatures reached at 
RHIC ($T \sim 200~{\rm MeV}$). At high temperatures, more phenomenological 
approaches come into play and ChPT provides a useful tool to control the low 
temperature limit of such phenomenological models.

In ref. \cite{martemy04}, the modifications of kaon properties in isotopically 
symmetric hot pion matter has been calculated. Starting from kaon in-medium 
masses and mean field potentials calculated in isotopically symmetric pion 
matter to one loop of chiral perturbation theory, the results have been 
extended to RHIC temperatures using experimental data on $\pi K$ scattering 
phase shifts. Kaon in-medium broadening and $\phi$ in-medium width were 
calculated as well. Here we are using the results of ref. \cite{martemy04} 
to investigate the role of the rescattering of secondary kaons inside the 
pion matter on the apparent dilepton branching ratio of the $\phi$ meson.

The paper is organised as follows. In section \ref{medium_mod} we briefly 
review the main results of kaon in-medium masses and broadening, mean field 
potential, as well as of $\phi$ meson in-medium width in an isotopically 
symmetric hot gas of pions investigated in ref. \cite{martemy04}. In section 
\ref{apparent_branch} we discuss the identification of $\phi$ mesons from 
both the dimuon and dikaon channels and calculate its apparent dilepton 
branching ratio in the framework of an evolutionary Bjorken scenario for the 
expanding fireball. Summary and conclusions are finally given in section 
\ref{summary}.

%%%%%%%%%%%%%%%%%%%%%%%%%%%%%%%%%%%%%%%%%%%%%%%%%%%%%%%%%%%%%%%%%%%%%%%%%%

\section{In-medium modification of kaons in pion matter}
\label{medium_mod}

We start with the $\phi$ meson in-medium width, which can be written in the 
form
\begin{equation}
\Gamma_{\phi \rightarrow K\bar{K}}^{*} = \eta~ 
\Gamma_{\phi \rightarrow K\bar{K}}^{{\rm vac}}~,
\end{equation}
where the $\eta$ function
\begin{eqnarray}
\eta &=& \frac{1}{\pi ^{2}} \int 
\frac{p^{*3}(M_{\phi}^{*},m_{1}^{*},m_{2}^{*})}{p^{*3}(M_{\phi},M_{K},M_{K})}~ 
\frac{M_{K}^{*}\Gamma_{K}^{*}dm_{1}^{*2}}{(m_{1}^{*2} - M_{K}^{*2})^{2} + 
(M_{K}^{*}\Gamma_{K}^{*})^{2}} \nonumber \\
&& \times \frac{M_{K}^{*}\Gamma_{K}^{*}dm_{2}^{*2}}{(m_{2}^{*2} - M_{K}^{*2})^{2} 
+ (M_{K}^{*} \Gamma_{K}^{*})^{2}}
\end{eqnarray}
contains the in-medium changes of the $\phi$ meson decay width. Here 
$p^{*3}(M_{\phi},M_{K},M_{K})$ and $p^{*3}(M_{\phi}^{*},M_{K}^{*},M_{K}^{*})$ 
are the kaon three-momenta in the $\phi$ meson rest frame for an in-vacuum 
and in-medium decay respectively. $\Gamma_{K}^{*}$ is the kaon collision 
width. The in-medium masses are $M_{\phi}^{*} = M_{\phi} - 2V_K$ and 
$M_{K}^{*} = M_{K} + \delta M_{K}$, where $V_K$ is the mean field potential 
and $\delta M_{K}$ is the mass shift for kaons in isotopically symmetric 
pion matter \cite{martemy04}.

The kaon collision width is given by
\begin{eqnarray}
\Gamma_{K}^{*} &=& \frac{1}{6M_{K}(2\pi)^{2}} \sum_{\ell} \int 
\left[ \left| A_{\ell}^{1/2}(s) \right|^{2} + 
2 \left| A_{\ell}^{3/2}(s) \right|^{2}\right]~ dn_{s\pi} \nonumber \\
&& \times \Phi_{2}(\sqrt{s},M_{\pi},M_{K})~,
\label{gammak}
\end{eqnarray}
where $A_{\ell}^{I}(s)$ are the on-shell partial wave projections of the 
$\pi K$ amplitudes with total isospin $I=1/2$ and $3/2$. The function 
$\Phi_2(\sqrt{s},M_{\pi},M_{K}) = \pi p^{*3}(\sqrt{s},M_{\pi},M_{K})/\sqrt{s}$ 
is the invariant $\pi K$ phase space with the c.m. momentum of the $\pi K$ 
system $p^{*}(\sqrt{s},M_{\pi},M_{K})$. The scalar pion density is defined by 
$dn_{s\pi} = dn_{v\pi}/(2E_{\pi})$ via the number densities of pions given by 
the Bose distribution function
\begin{equation}
dn_{v\pi} = \frac{d^{3}p_{\pi}}{(2\pi)^{3}}
\left[ \exp\left( \frac{E_{\pi}-\mu _{\pi }}{T}\right) - 1\right]^{-1}~,
\end{equation}
where $\mu_{\pi ^{+}} = -\mu_{\pi^{-}}$ is the $\pi^{+}$ chemical potential 
and $\mu_{\pi^{0}} = 0$.

\begin{figure*}[htb]
\resizebox{1.0\textwidth}{!}{\includegraphics{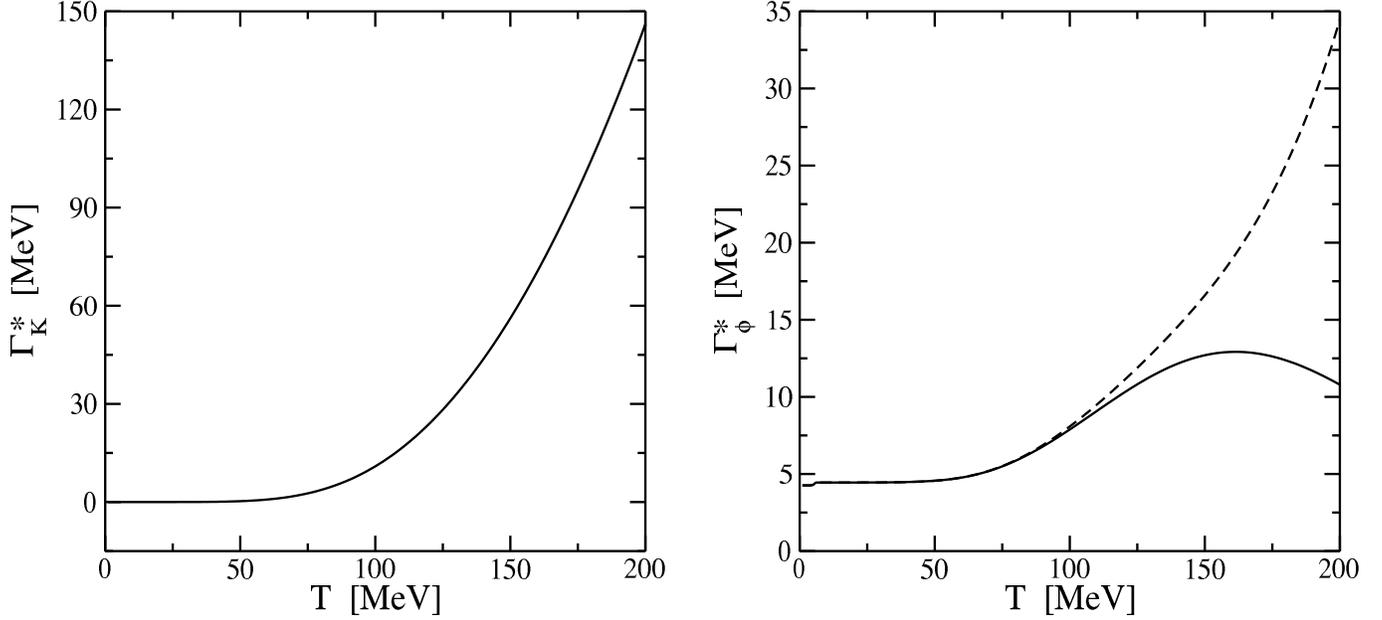}}
\caption{Kaon collision width $\Gamma_{K}^{*}$ (left) and in-medium $\phi$ 
meson width $\Gamma_{\phi \rightarrow K\bar{K}}^{*}$ (right, solid line) 
versus temperature $T$ in isotopically symmetric pion matter. The dashed 
curve in the right panel corresponds to a phenomenologically increased 
$\phi$ width $\Gamma_{\phi}^{tot}=\Gamma_{\phi}^{*}+\Gamma_{\phi+\rho}^{coll}$, 
in which an additional collisional broadening by $\rho$ mesons is taken into 
account (see the detailed discussion in section \ref{apparent_branch}).
}
\label{fig1}
\end{figure*}

The on-shell $\pi K$ amplitudes have been calculated in ChPT to the order 
$p^{4}$ by several authors (see {\it e.g.} \cite{pik} and references therein). 
To lowest order $p^{2}$, the off-shell $\pi K$ amplitudes are given in ref. 
\cite{Weinberg}. 
The self energy operator $\Sigma (M_{K}^2,M_K)$, mass-shift $\delta M_{K}$ 
and mean field potential $V_{K}$ for kaons in an isotopically symmetric pion 
gas have been calculated to leading  order in pion density 
in chiral perturbation theory from the $\pi K$ forward 
scattering amplitudes determined at the one loop level. The
amplitudes were expressed in terms 
of the $s$- and $p$-wave $\pi K$ scattering lengths and the $s$-wave 
effective ranges. The derived expressions, valid at $T\lesssim M_{\pi }$, 
can be extended via a more phenomenological approach to typical chemical 
freeze-out temperatures at RHIC ($T\sim 170$ MeV \cite{thermo}) by rewriting 
the $s$-wave parts of the amplitudes $A^{I}$ in terms of phase shifts and 
by assuming the behaviour of the $p$-wave to be fixed by the $a_{1}^{I}$ 
scattering length and the resonance $K^{*}$. The resulting amplitudes satisfy 
unitarity. The experimental $\pi K$ scattering phases are then well 
reproduced; the low-temperature limit matches smoothly with one-loop ChPT. 
For the details, we refer to ref. \cite{martemy04}. 
The resulting kaon collision width and in-medium $\phi$ meson width are shown
in fig. \ref{fig1} as a function of temperature in the left and right panel 
respectively. 

%%%%%%%%%%%%%%%%%%%%%%%%%%%%%%%%%%%%%%%%%%%%%%%%%%%%%%%%%%%%%

\section{Apparent dilepton branching ratio of the $\phi$ meson}
\label{apparent_branch}

Since the $\phi$ meson is unstable, it can only be detected from its decay 
products of either the kaon-antikaon pair or the lepton pair. Dileptons from 
$\phi$ decays leave the pion matter essentially undistorted by final state 
interactions, whereas the secondary kaons rescatter with the medium and can 
contribute to the experimental background. We investigate whether this can 
result in an increase of the apparent dilepton branching ratio or not.

Let us assume that $\phi$ mesons leave the heavy ion reaction at time $\tau$. 
In the simple case of \emph{constant widths}, as already pointed out in 
ref. \cite{martemy04}, the probability for the kaons to leave the reaction 
zone without rescattering is given by
\begin{equation}
w \sim \int_{0}^{\tau} 
\exp\left(-2~ \Gamma _{K}^{*}~ (\tau - t)\right)~ 
\exp\left(-\Gamma_{\phi}^{*}~ t\right)~\Gamma_{\phi}~ dt~.
\label{w}
\end{equation}
The first factor in this expression is the probability for two kaons to escape 
from the reaction zone without rescattering. The second factor is the survival 
probability of the $\phi$ meson at time $t$. The last factor $\Gamma_{\phi}~dt$ 
is the probability to decay into the kaon pair during $dt$. Notice that 
$\Gamma_{\phi}^{*}~dt$ has the meaning of a decay probability into the kaons 
which rescatter with pions from the surrounding medium. Selecting kaon pairs 
with invariant masses of the $\phi$ meson which suffered no rescatterings, 
the number of $\phi$ mesons observed in the $K\bar{K}$ channel equals 
\begin{eqnarray}
N_{K\bar{K}} \sim \exp\left(-\Gamma_{\phi}^{*}~ \tau\right) &+& 
\frac{\Gamma_{\phi}}{2~ \Gamma_{K}^{*} - \Gamma_{\phi}^{*}} 
\left[ \exp\left(-\Gamma_{\phi}^{*}~ \tau\right) \right. \nonumber \\
&& \left. - \exp\left(-2~ \Gamma_{K}^{*}~ \tau\right)\right]~.
\label{NKK}
\end{eqnarray}
The first term arises due to the vacuum decays, whereas the second term is 
given by eq. (\ref{w}). The number of $\phi$ mesons observed in the dilepton 
channel equals 
\begin{equation}
N_{e^{+}e^{-}} \sim B \exp\left(-\Gamma_{\phi}^{*}~ \tau\right) + 
B^{^{*}} \left[ 1 - \exp\left(-\Gamma_{\phi}^{*}~ \tau\right)\right]~,
\label{Nee}
\end{equation}
where $B$ and $B^{*}$ are the vacuum and in-medium dilepton branching ratios 
of the $\phi$ meson respectively. We assume that the dilepton channel suffers 
no in-medium modifications, 
{\it i.e.} $B^{*}/B = \Gamma_{\phi}/\Gamma_{\phi}^{*}$. 
Notice that the second term of eq. (\ref{Nee}) can be also written as 
\begin{eqnarray}
\nonumber
B^{^{*}} \left[ 1 - \exp\left(-\Gamma_{\phi}^{*}~ \tau\right)\right] &=& 
\int_{0}^{\tau} 
\exp\left(-\Gamma_{\phi}^{*}~ t\right)~ \Gamma_{\phi\rightarrow e^{+}e^{-}}~ dt 
\nonumber \\
&=& \int_{0}^{\tau} 
\exp\left(-\Gamma_{\phi}^{*}~ t\right)~ B~ \Gamma_{\phi}~ dt~,
\end{eqnarray}
whose interpretation is straightforward, being $\exp(-\Gamma_{\phi}^{*}~ t)$ 
the survival probability of the $\phi$ meson at time $t$ and 
$\Gamma_{\phi\rightarrow e^{+}e^{-}}~ dt$ its probability to decay into a 
dilepton pair during $dt$.

Turning now to \emph{temperature dependent}, and therefore 
\emph{time dependent widths}, the previous relations should be 
modified in the following way:
\begin{eqnarray}
\nonumber
\exp\left(-2~ \Gamma_{K}^{*}~ (\tau - t)\right)\quad &\Longrightarrow& 
\quad 
\exp\left(-2\!\int_{t}^{\tau}\!\!\Gamma_{K}^{*}(t')~ dt'\right) 
\nonumber \\
\exp\left(-\Gamma_{\phi}^{*}~ t\right) \quad &\Longrightarrow& \quad 
\exp\left(-\int_{0}^{t}\!\!\Gamma_{\phi}^{*}(t')~ dt'\right) 
\nonumber \\
\exp\left(-\Gamma_{\phi}^{*}\tau\right) \quad &\Longrightarrow& \quad 
\exp\left(-\int_{0}^{\tau}\!\!\Gamma_{\phi}^{*}(t')~ dt'\right) 
\nonumber
\end{eqnarray}
so that the number of $\phi$ mesons decaying into $K\bar{K}$ pairs 
is determined by
\begin{eqnarray}
N_{K\bar{K}} &\sim& \exp\left(-\!\int_{0}^{\tau}\!\!\!
\Gamma_{\phi}^{*}(t')~dt'\right) +\! \int_{0}^{\tau}\!\!\!
\exp\left(-2\!\int_{t}^{\tau}\!\!\!
\Gamma_{K}^{*}(t')~dt'\right) \nonumber \\
&& \times \exp\left(-\!\int_{0}^{t}\!\!
\Gamma_{\phi}^{*}(t')~dt'\right)~\Gamma_{\phi}~dt~,
\label{NKK2}
\end{eqnarray}
and those decaying into dileptons by
\begin{eqnarray}
N_{e^{+}e^{-}} &\sim& 
B \exp\left(-\int_{0}^{\tau}\!\!\Gamma_{\phi}^{*}(t')~dt'\right) \nonumber \\
&& +~ B\!\int_{0}^{\tau}\!\!\exp\left(-\int_{0}^{t}\!\!
\Gamma_{\phi}^{*}(t')~ dt'\right) 
\Gamma_{\phi }~dt~.
\label{Nee2}
\end{eqnarray}
Finally, the apparent dilepton branching ratio with respect to the vacuum 
branching ratio is then simply given by the ratio of both quantities 
(\ref{NKK2}) and (\ref{Nee2}), {\it i.e.}
\begin{equation}
\frac{B^{app}}{B} = \frac{N_{e^{+}e^{-}}}{N_{K\bar{K}}}~.
\label{Bapp}
\end{equation}

In central relativistic heavy ion collisions the distribution of matter is 
approximatively uniform in rapidity, at least in the mid-rapidity region, 
and the geometry of the collision can be considered as cylindrically 
symmetric \cite{Bjorken83}. 
In the framework of a Bjorken scenario, extensively used to describe the 
space-time evolution of ultra-relativistic heavy ion collisions, the time 
and temperature of the evolving system are related according to 
\cite{Bjorken83} by
\begin{equation}
\nonumber
\frac{T(t)}{T_0} = \left( \frac{t_0}{t} \right)^{c_s^2}~,
\label{bjorkenT0t0}
\end{equation}
where $c_s^2$ is the sound velocity squared for a baryon less gas, and 
$T_0$ is the initial temperature of the system that is assumed to be 
fully thermalized after the time $t_0$. Because our investigations are 
restricted to matter created in heavy ion reactions after hadronization, 
we have applied eq. (\ref{bjorkenT0t0}) to describe the temporal evolution 
of the hadronized system reinterpreting $t_0$ and $T_0$ by the hadronization 
time $t_h$ and hadronization temperature $T_h$ respectively. 
For an ideal relativistic gas, the energy density and the pressure are 
related by $\epsilon = 3 p$, therefore one finally has 
\begin{equation}
\frac{1}{c_s^2} = \frac{d\epsilon}{dp} = 3 \qquad {\rm and} \qquad 
\frac{T(t)}{T_h} = \left( \frac{t}{t_h}\right)^{-1/3}~.
\label{bjorkenThth}
\end{equation}

\begin{figure}[htb]
\resizebox{1.0\columnwidth}{!}{\includegraphics{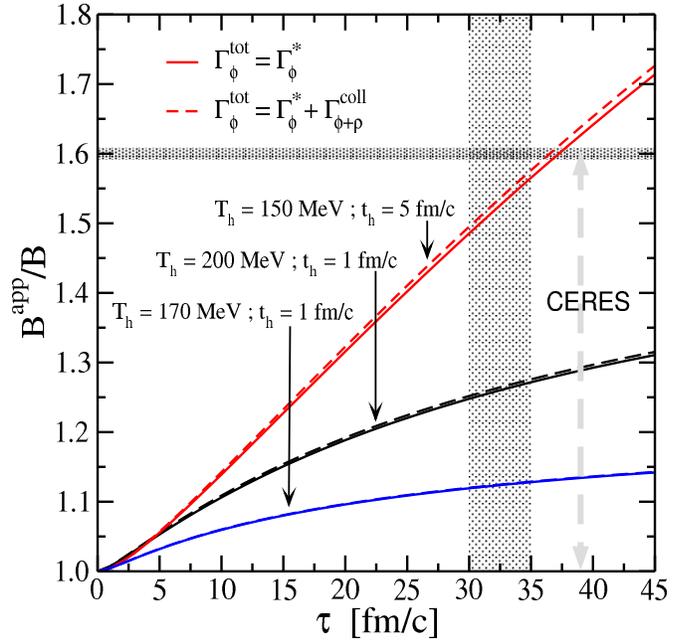}}
\caption{Ratio between the apparent and vacuum dilepton branchings 
$B^{app}/B$ versus the $\phi$ meson's escape time $\tau$ (solid curves). 
The dashed curves correspond to the case in which an additional collisional 
broadening in the phi meson width is taken into account (see right panel of 
fig. \ref{fig1}). The vertically hatched area corresponds to a region around 
a $\tau$ value of $\approx 32~{\rm fm/c}$ predicted by transport calculations 
for RHIC conditions. The horizontally hatched band at $B^{app}/B = 1.6$ 
indicates the upper limit of the enhancement factor according to the 
observations of the CERES Collaboration at SPS (see discussion below).
}
\label{fig2}
\end{figure}

In section \ref{medium_mod}, expressions for the kaon collision width and 
in-medium $\phi$ meson width as functions of the temperature were given. 
The apparent dilepton branching ratio can be evaluated from eq. (\ref{Bapp}) 
using eqs. (\ref{NKK2}) and (\ref{Nee2}) by performing the $t \rightarrow T$ 
variable replacement according to the transformation law (\ref{bjorkenThth}). 
In addition to the onset of the $T(t)$ evolution, the parameters $T_h$ and 
in particular $t_h$ determine mainly the slope of the temperature decrease 
with ongoing expansion. Thus, both parameters may strongly influence the 
final results in the following way: the larger $T_h$ and $t_h$ are, {\it i.e.} 
the slowlier the hadronic phase cools down, the stronger the in-medium 
effects are. The resulting ratios $B^{app}/B$ for three different sets of 
these parameters are shown in fig. \ref{fig2} as a function of the phi 
mesons's escape time $\tau$. 
First of all, we chose in our model calculation 
an initial temperature of $T_h = 200~{\rm MeV}$ and a corresponding time of 
$t_h = 1~{\rm fm/c}$ leading to a fast cooling behaviour of the expanding 
fireball. In this quasi extreme scenario, we find that the phi meson yield 
from the dilepton channel is only slightly higher by a factor of about $1.25$ 
than that from the $K\bar{K}$ channel -- at least in the physically 
interesting $\tau$ region. According to transport calculations, the term 
$\exp(-\tau \Gamma_{\phi})$ is of the order $\sim 1/2$ at RHIC energies 
\cite{bravina03}, {\it i.e.} $\tau \sim 32~{\rm fm/c}$. Therefore, the 
expected region of physical interest is depicted in fig. \ref{fig2} as 
vertically hatched area. 
The $B^{app}/B$ ratio from a model calculation with practically the same fast 
cooling, but a smaller and therefore more realistic hadronization temperature 
of $T_h = 170~{\rm MeV}$ according to transition temperatures obtained from 
2-flavour lattice QCD \cite{Karsch:2003jg} is even essentially smaller, 
because the interval of the temperature integration is shrinked. 
The consequences of a larger hadronization time are also depicted in 
fig. \ref{fig2}. With a value of $t_h = 5~{\rm fm/c}$, the temperature of 
the expanding hadronic fireball decreases much slowlier compared to the 
previous parametrizations. In this case, the in-medium effects influence the 
ratio between the apparent and the vacuum dilepton branchings $B^{app}/B$ in 
a much stronger manner, that overcompensates the (converse) effect of an even 
smaller hadronization temperature $T_h = 150~{\rm MeV}$ according to transition 
temperatures obtained from 3-flavour lattice QCD \cite{Karsch:2003jg}. 
The enhancement factor in the latter scenario turns out to be about $1.5-1.6$ 
in the region around a $\tau$ value of $\approx 32~{\rm fm/c}$.

Nevertheless, all these values are in qualitative agreement with the 
observations of the CERES Collaboration at SPS energies, who excludes an 
enhancement factor larger than $1.6$ at 95\% CL\footnote{The $\phi$ yields in 
the leptonic and the hadronic channel published in ref. \cite{Adamova:2005jr} 
result in a ratio 
$B^{app}/B$ of $1.0 \pm 0.31~{\rm (stat)} \pm 0.28~{\rm (syst)}$, 
{\it i.e.} a maximum value of about $1.6$ (upper limit).}, and contrasts 
the strong enhancement factor of $2-4$ resulting from the comparison between 
the yields measured for the $\mu^+\mu^-$ and $K^+K^-$ channels by the NA50 
and NA49 Collaborations respectively. 
Well, a large enhancement of that magnitude is still not totally excluded 
within the simplified view of a Bjorken scenario, that assumes only a 
longitudinally expanding medium. We want to emphasise once more that the 
question, how fast the temperature of the hadronic fireball decreases, seems 
to be very crucial. A more realistic description taking additionally a 
transversal expansion into account might promote a fast cooling scenario, 
that tends to result in not so strong enhancement factors. 
The exclusion of a large enhancement due to in-medium effects and kaon 
collisional broadening in heavy ion collisions at RHIC energies results also 
from multi-phase transport calculations. In particular, the production of 
$\phi$ mesons reconstructed from dikaon and dilepton decays has been 
investigated in ref. \cite{Pal:2002aw} for central Au+Au collisions at a 
RHIC energy of $\sqrt{s_{\rm NN}} = 130~{\rm GeV}$. There have been included 
density dependent in-medium modifications of kaon masses derived from a chiral 
effective Lagrangian and a linear decrease of the $\phi$ mass according to a 
scaling law $m_{\phi}^{*} \approx m_{\phi}(1 - \alpha\rho/\rho_0)$, with 
$\alpha=0.0255$ and the normal nuclear matter density $\rho_0$. It was found 
that the $\phi$ meson yield from the dilepton channel is about $1.5$ times 
larger than that from the $K^+K^-$ channel in the absence of additional 
$\phi$ meson production due to the formation of an initial partonic stage.

The idea of a $\phi$ meson which decouples easily from the hadronic fireball 
is widely diffused, as a consequence of the small cross section for its 
scattering with non-strange hadrons. However, available calculations give 
contradictory results \cite{Ko:1993id,Haglin:1994xu,Smith:1997xu,koch02}. 
Whereas some of them support this idea 
\cite{Ko:1993id,Haglin:1994xu,Smith:1997xu}, a calculation based on a Hidden 
Local Symmetry Lagrangian shows that the collision rates of the $\phi$ meson 
in a hot hadronic matter of pseudo-scalar ($\pi$, $K$) and vector mesons 
($\rho$, $\omega$, $K^*$, $\phi$) is rather large \cite{koch02} and dominated 
at temperatures between 150 MeV and 200 MeV by the $K^*$ followed by $K$ and 
$\rho$ contributions, while the contributions from $\pi$, $\omega$ and $\phi$ 
are smaller. 
The corresponding $\phi$ mean free path at temperatures above 170 MeV is 
between 2.4 fm and 1 fm, and therefore considerably smaller than the typical 
size of the hadronic system created in heavy ion collisions 
($\sim 10-15~{\rm fm}$). This would imply that most $\phi$ mesons produced 
after the hadronization would not leave the hadronic system without 
scattering. It was found, that reactions changing the number of $\phi$ 
mesons account for more than 80\% of the total collision rate, so that at the 
early stage of the expansion, when the effect of the decay is negligible, the 
collision rate is mainly responsible for the decrease of the $\phi$ number 
with respect to its value at the hadronization.

Since a pion gas is only the first approximation to the hadronic final state 
of a relativistic heavy ion collision one can estimate the stability of the 
present investigations by the inclusion of the 
$\phi + \rho \rightarrow K\bar{K}$ channel, {\it i.e.} the next important 
channel which gives -- according to the results in ref. \cite{koch02} -- rise 
to in-medium modifications of the $\phi \rightarrow K\bar{K}$ decay. However, 
in contrast to the influence of the pion gas \cite{martemy04} we consider the 
contributions from the $\rho$ mesons in a phenomenological way: 
following ref. \cite{koch02} the collisional width $\Gamma_{\phi + \rho}^{coll}$ 
due to the scattering of $\phi$'s on $\rho$ mesons is added to the in-medium 
width $\Gamma_{\phi}^{*}$ from the interaction with the pions. To do so 
we extrapolated the results of ref. \cite{koch02} imposing the boundary 
condition $\Gamma_{\phi+\rho}^{coll} = 0$ at $T = 0$ as well as a fast decrease 
to zero for $T \rightarrow 0$ and calculated the apparent dilepton branching 
ratio using this time a total $\phi$ meson width equal to 
$\Gamma_{\phi}^{tot} = \Gamma_{\phi}^{*} + \Gamma_{\phi + \rho}^{coll}$. 
Fig. \ref{fig1} shows the obtained total $\phi$ meson width 
$\Gamma_{\phi}^{tot}$ as a function of temperature. Obviously in a very 
high but narrow temperature range, the $\phi + \rho$ collision rate dominates 
on the in-medium $\phi$ decay width. When the system cools down, the 
additional rate drops fast and becomes negligible compared to 
$\Gamma_{\phi}^{*}$ in the lower temperature domain.

Fig. \ref{fig2} depicts also the corresponding ratios $B^{app}/B$ with 
comparison to their values obtained when the collisional broadening due to 
$\rho$ mesons is neglected. As one can easily see, the introduction of a 
collisional broadening does not significantly influence the apparent dilepton 
branching ratio. The increase of the total $\phi$ width reduces simultaneously 
the phi decays into dileptons as well as the number of $\phi$ mesons 
reconstructed via the $K\bar{K}$ decay channel. The two reduction factors 
turn out to be comparable in magnitude, so that at the end the ratio 
$B^{app}/B$ is only very slightly additionally enhanced. This fact is retained 
for all of the three parametrizations being under consideration. Furthermore, 
this fact demonstrates that for the present purpose the fireball expansion 
after hadronization can safely be described by an expanding pion gas. 
The reason is simply that the integration over the entire space-time volume 
of the expanding system is dominated by the lower temperatures where the 
heavier mesons are suppressed by their Boltzmann factors.

%%%%%%%%%%%%%%%%%%%%%%%%%%%%%%%%%%%%%%%%%%%%%%%%%%%%%%%%%%%%%%%%%%%%%%%%%

\section{Summary and conclusions}
\label{summary}

We have investigated the role of the rescattering of secondary kaons inside 
pionic matter on the apparent dilepton branching ratio of the $\phi$ meson 
in heavy ion collisions at RHIC energies, taking into account in-medium mass 
modifications and broadening of kaons and $\phi$ mesons in an isotopically 
symmetric hot pion gas. We have found within the framework of a Bjorken 
scenario for the temporal evolution of the expanding hadronic fireball that 
the phi meson yield reconstructed from the dilepton channel is only by a 
factor between about $1.1$ and $1.6$ higher than that reconstructed from 
the kaon-antikaon channel. We have demonstrated that the magnitude of this 
enhancement depends also on the question, how slowly the temperature of the 
pionic medium decreases. Nevertheless, for reasonable parametrizations of 
the cooling scenario, our results agree quantitatively well with recent 
observations of the CERES Collaboration at CERN SPS energies. 
The inclusion of further in-medium effects through the addition of a possible 
collisional broadening in the $\phi$ meson in-medium width does not change 
considerably our results. For heavy ion collisions at RHIC energies, only 
preliminary results on the $\phi$ meson yield measured via the two decay 
channels are available up to now. The errors in the dilepton channel are 
unfortunately still too large for a definitive statement. Future data from 
RHIC will help bring more insight into the meson propagation in a dense and 
hot medium.

%%%%%%%%%%%%%%%%%%%%%%%%%%%%%%%%%%%%%%%%%%%%%%%%%%%%%%%%%%%%%%%%%%%%%%%%%

\section*{Acknowledgements}

This work was supported by the European Graduate School 
Basel--Graz--T{\"u}bingen and by the Bundesministerium f{\"u}r Bildung und 
Forschung (BMBF) under contract 06T{\"U}202.

%%%%%%%%%%%%%%%%%%%%%%%%%%%%%%%%%%%%%%%%%%%%%%%%%%%%%%%%%%%%%%%%%%%%%%%%%

%%%%%%%%%%%%%%%%%%%%%%%%%%%%%%%%%%%%%%%%%%%%%%%%%%%%%%%%%%%%%%

\end{document}